# The Puzzling Long GRB 191019A: Evidence for Kilonova Light

G. Stratta[1,2,3,4], A. M. Nicuesa Guelbenzu[5], S. Klose[5], A. Rossi[2], P. Singh[1,2], E. Palazzi[2], C. Guidorzi[2,6,7], A. Camisasca[8], S. Bernuzzi[9], A. Rau[10], M. Bulla[6,7,11], F. Ragosta[12,13], E. Maiorano[2], and D. Paris[14]
[1] Institut für Theoretische Physik, Goethe Universität, Max-von-Laue-Str. 1, D-60438 Frankfurt am Main, Germany
[2] INAF—Osservatorio di Astrofisica e Scienza dello Spazio, Via Piero Gobetti 93/3, 40129 Bologna, Italy
[3] Istituto di Astrofisica e Planetologia Spaziali, Via Fosso del Cavaliere 100, I-00133 Roma, Italy
[4] Istituto Nazionale di Fisica Nucleare—Roma 1, Piazzale Aldo Moro 2, I-00185 Roma, Italy
[5] Thüringer Landessternwarte Tautenburg, Sternwarte 5, 07778 Tautenburg, Germany
[6] Department of Physics and Earth Science, University of Ferrara, Via Saragat 1, 44122 Ferrara, Italy
[7] INFN—Sezione di Ferrara, Via Saragat 1, 44122 Ferrara, Italy
[8] Astronomical Observatory of the Autonomous Region of the Aosta Valley (OAVdA), Loc. Lignan 39, I-11020, Nus (Aosta Valley), Italy
[9] Theoretisch-Physikalisches Institut, Friedrich-Schiller-Universität Jena, 07743 Jena, Germany
[10] Max-Planck-Institut für Extraterrestrische Physik, Giessenbachstraße 1, 85748 Garching, Germany
[11] INAF, Osservatorio Astronomico d'Abruzzo, Via Mentore Maggini snc, 64100 Teramo, Italy
[12] Dipartimento di Fisica "Ettore Pancini," Università di Napoli Federico II, Via Cinthia 9, 80126 Naples, Italy
[13] INAF—Osservatorio Astronomico di Capodimonte, Via Moiariello 16, I-80131 Naples, Italy
[14] INAF—Osservatorio Astronomico di Roma, Via Frascati 33, 00040 Monte Porzio Catone, Italy
Received 2024 August 10; revised 2024 November 13; accepted 2024 December 4; published 2025 January 23

## Abstract

GRB 191019A was a long gamma-ray burst (GRB) lasting ~65 s and, as such, originally thought to be linked to a core-collapse supernova. However, even though follow-up observations identified the optical counterpart close to the bright nucleus of a nearby ancient galaxy ($z = 0.248$), no associated supernova was found. This led to the suggestion that the burst was caused by the merger of two compact stellar objects, likely in a dense circumnuclear environment. By using a recently developed diagnostic tool based on prompt emission temporal properties, we noticed that GRB 191019A falls among those long GRBs which are associated with compact mergers and with evidence of kilonova light. We thus reanalyzed unpublished GROND multicolor ($g'r'i'z'JHK_s$) data obtained between 0.4 and 15 days posttrigger. Image subtraction confirmed the optical counterpart in all four optical bands, with GROND tracking its fading until 1.5 days postburst. Incorporating publicly available Swift-XRT data, a joint fit of an afterglow plus a kilonova model revealed a better match than an afterglow-only scenario. The resulting kilonova properties resemble those of AT2017gfo associated with the binary neutron star merger GW170817, with a total ejected mass of ~0.06 $M_\odot$. Contrary to previous findings inferring a high-density circumburst environment ($n_0 \sim 10^{7-8}$ cm$^{-3}$), our analysis finds standard conditions ($n_0 \sim 1$ cm$^{-3}$), suggesting the long duration of GRB 191019A was intrinsic rather than due to jet interaction with a dense external medium.

*Unified Astronomy Thesaurus concepts:* Gamma-ray bursts (629); Compact objects (288)

## 1. Introduction

The empirical classification of gamma-ray bursts (GRBs) into two classes according to the bimodal burst duration and hardness distribution (see, e.g., C. Kouveliotou et al. 1993), has been interpreted as evidence of two different progenitor channels. Indeed, it was found that the vast majority of sufficiently nearby long GRBs are found to be spatially and temporally coincident with core-collapse (CC) supernovae (SNe) and hosted in star-forming galaxies. On the other hand, short GRBs lack any associated SN and are typically located in the outer regions of early-type galaxies, suggesting a progenitor belonging to an old stellar population, formally consistent with compact binary mergers. This picture was confirmed by the association of the short GRB 170817A with the gravitational-wave event GW170817 that was compatible with a binary neutron star (NS–NS) merger (e.g., B. P. Abbott et al. 2017). From the very same event and thanks to its spatial proximity (40 Mpc), the first robust evidence of a kilonova (AT2017gfo) was obtained (e.g., D. A. Coulter et al. 2017), confirming the predicted thermal emission from matter released and heated during an NS–NS merger (e.g., L.-X. Li & B. Paczyński 1998). In the last decade, kilonova signatures were observed in a number of past short GRBs that were sufficiently nearby, as for instance GRB 130603B ($z = 0.3565$; N. R. Tanvir et al. 2013), GRB 150101B ($z = 0.134$; E. Troja et al. 2018), and GRB 160821B ($z = 0.1613$; E. Troja et al. 2019), though with much lower significance than AT2017gfo due to the larger distances of these events (see, e.g., E. Troja 2023 for a review).

In recent years, the standard long- and short-GRB progenitor paradigm has been blurred by mounting evidence of the existence of long GRBs with no associated CC-SN but with evidence of a kilonova component, more consistent with a compact binary merger progenitor. Two striking examples are the long GRB 211211A at redshift $z = 0.076$ (350 Mpc; e.g., E. Troja et al. 2018; A. Mei et al. 2022; J. C. Rastinejad et al. 2022; J. Yang et al. 2022; B. P. Gompertz et al. 2023) and GRB 230307A at $z = 0.0646$ (300 Mpc; e.g., A. J. Levan et al. 2023), for which a kilonova was clearly identified in the optical afterglow. In addition, evidence of short GRBs showing massive-star collapse properties were also found (e.g., T. Ahumada et al. 2021; A. Rossi et al. 2022), which further contributed to dismantle the standard GRB classification scheme. These results suggest that some of the past GRB progenitor classifications based on the burst duration and hardness only have to be revisited. This holds in particular for GRB 191019A.

GRB 191019A was discovered (K. K. Simpson et al. 2019) with the Burst Alert Telescope (BAT; S. D. Barthelmy et al. 2005) on board the Neil Gehrels Swift Observatory (Swift; N. Gehrels et al. 2004). It showed a complex burst light curve with multiple peaks with a duration $T_{90} = 64.6 \pm 4.5$ s. Its 15–150 keV spectrum could be well modeled with a power law with a photon index of $2.25 \pm 0.05$ and a fluence of $10^{-5}$ erg cm$^{-2}$ (H. A. Krimm et al. 2019) putting this source among the soft, long GRBs. Follow-up observations with the Swift X-ray telescope (XRT; D. N. Burrows et al. 2005) robustly identified an uncatalogued X-ray source (P. A. Evans et al. 2019). Rapid follow-up observations with the Zeiss-1000 1 m telescope of Tien Shan Astronomical Observatory revealed only one optical source within the XRT error circle with $R = 18.37 \pm 0.03$ mag at $T_0 + 14$ minutes (I. Reva et al. 2019). The position of this source was coincident with a cataloged object quoted to be fainter than the observed one, leading to the interpretation this is the host galaxy of GRB 191019A with the afterglow on top of it (I. Reva et al. 2019). The Zadko telescope (D. M. Coward et al. 2017) observed the GRB 191019A field at similar epochs and confirmed the object reported by I. Reva et al. (2019) with $R = 18.92$ mag, with no significant flux variation within 1.5 hr (B. Gendre et al. 2019). Similar claims of a lack of flux variations were provided by other teams performing early epoch observations (J. P. U. Fynbo et al. 2019, A. Nicuesa Guelbenzu 2019, D. A. Perley et al. 2019a, Z. P. Zhu et al. 2019). However, late observations performed with the Nordic Optical Telescope (NOT) at $T_0 + 3.25$ days were analyzed with image subtraction methods and revealed that the source was fading, confirming the optical transient plus host identification (D. A. Perley et al. 2019b).

Spectroscopic analysis of the host galaxy showed several absorption lines at redshift $z = 0.248$ and a spectral energy distribution (SED) consistent with a galaxy dominated by an old stellar population (>1 Gyr), with a star formation rate (SFR) of $0.06 \pm 0.03\ M_\odot$ yr$^{-1}$, a stellar mass of $3 \times 10^{10}\ M_\odot$, and small dust extinction of $A_V = 0.19 \pm 0.08$ mag (A. J. Levan et al. 2023). The measured SFR is at odds with the typical high values inferred in long-GRB hosts (but see A. Rossi et al. 2014). Further hints of anomalies for the long GRB 191019A came from the absence of an associated SN, despite the low redshift. Deep limits in $g$, $r$, and $z$ obtained between 2 and 73 days with the NOT and optical imaging with the Hubble Space Telescope (HST) at 30 and 184 days, put strong constraints on any SN emission up to >20 times less luminous than SN 1998bw (A. J. Levan et al. 2023). All these properties are at odds with a massive-star origin of GRB 191019A, while formally consistent with a compact object binary merger progenitor. Though, the deep limits obtained with the NOT Telescope at >2 days could not confirm the presence of early kilonova emission (A. J. Levan et al. 2023).

An interesting feature characterizing GRB 191019A is its projected distance from the host-galaxy center of $\lesssim 100$ pc, the closest distance among all known short GRBs to their corresponding galactic nucleus (A. J. Levan et al. 2023). Indeed, compact binary progenitors formed in stellar binary systems are thought to have large kick velocities acquired during the CC-SN phase of the binary components, and for this reason short GRBs are typically found in the outskirts of their host galaxies, with large offsets from the center (e.g., E. Berger 2014; B. O'Connor et al. 2022; W.-f. Fong et al. 2022). As noted by A. J. Levan et al. (2023), the proximity of GRB 191019A to the host-galaxy center suggests that the possible progenitor compact binary system could have formed through dynamical encounters, which are thought to be favored in the dense gaseous environment of supermassive black hole surrounding disks through kinetic energy dissipation ("gas-capture" binary formation channel; H. Tagawa et al. 2020). An exciting consequence of this scenario is that dense environments could also alter the prompt emission properties of a short GRB, making it longer and softer (D. Lazzati et al. 2023). This possibility was explored for GRB 191019A and it was found compatible with an environmental density of the order of $10^7$–$10^8$ cm$^{-3}$ (D. Lazzati et al. 2023). Starting from D. Lazzati et al. (2023)'s conclusions, S.-N. Wang et al. (2024) investigated the interaction of a possible kilonova ejecta from GRB 191019A with a dense circumstellar medium (CSM) that could be detected years after the merger. Their model predicts that in a very dense environment, the smaller the kilonova ejected mass, the fainter is the radioactively powered luminosity but the higher is the contribution from kilonova–CSM interaction at late times. No evidence of kilonova ejecta–CSM interaction was found for GRB 191019A so far, and a possible detection in the future would require an ejected mass less than $2 \times 10^{-5}\ M_\odot$ (S.-N. Wang et al. 2024).

Here we further investigate the properties of GRB 1901910A and the nature of the optical transient that followed the burst, with the goal to find arguments in favor or against its possible compact merger origin. We use the recently developed GRB prompt emission minimum variability timescale (MVT) criterion (A. E. Camisasca et al. 2023) to explore the properties of the burst, and perform a complete reanalysis of the light curve of the optical/X-ray transient with particular emphasis on so far not analyzed data from the GROND instrument mounted on the ESO/MPG 2.2 m telescope at La Silla, Chile (J. Greiner et al. 2008), starting 10.3 hr postburst (A. Nicuesa Guelbenzu 2019). We use the sophisticated Nuclear Multimessenger Astronomy (NMMA; T. Dietrich et al. 2020; P. T. H. Pang et al. 2023) software package, which allows for afterglow and kilonova joint Bayesian inference, to explore if the multicolor light curve of the optical transient can be understood as powered by afterglow emission only or if a kilonova component was present too.

Throughout this paper, we adopt a flat cosmological model with $H_0 = 67.4$ km s$^{-1}$ Mpc$^{-1}$, $\Omega_M = 0.315$, and $\Omega_\Lambda = 0.685$ (Planck Collaboration et al. 2020). For these parameters a redshift of $z = 0.248$ (J. P. U. Fynbo et al. 2019) corresponds to a luminosity distance of $d_L = 1.29$ Gpc, 1″ corresponds to a 4.03 kpc projected distance, and the distance modulus is $m - M = 40.55$ mag. The Milky Way reddening $E(B - V)$ along the line of sight toward the source is between 0.03 and 0.04 mag (D. J. Schlegel et al. 1998; E. F. Schlafly & D. P. Finkbeiner 2011). We present the observations and data reduction in Section 2, and our results in Section 3. In our data analysis we corrected the apparent magnitudes for Galactic reddening and adopted a host-galaxy visual extinction along the line of sight of $A_V^{\rm host} = 0.06$ mag (A. J. Levan et al. 2023). Discussion and conclusions are presented in Section 4.

## 2. Observations and Data Reduction

### 2.1. GROND Early-time Observations

With a duration of about 65 s, GRB 191019A was a classical long GRB. At a redshift of $z = 0.248$, a rising SN 1998bw

**Table 1**
Measured AB Magnitudes and Upper Limits of the Transient

| Epoch | dt (days) | g′ | r′ | i′ (mag) | z′ | J | H | $K_s$ |
|---|---|---|---|---|---|---|---|---|
| (1) | (2) | (3) | (4) | (5) | (6) | (7) | (8) | (9) |
| 1a | 0.4413 | 23.16 ± 0.11 | 22.99 ± 0.10 | 22.78 ± 0.19 | 22.20 ± 0.23 | >21.4 | >20.8 | >20.1 |
| 1b | 0.5889 | 23.57 ± 0.40 | 23.17 ± 0.27 | >22.8 | >22.4 | >21.1 | >20.5 | >19.9 |
| 2 | 1.5006 | 23.96 ± 0.15 | 23.61 ± 0.15 | 23.46 ± 0.28 | >22.8 | >21.4 | >20.7 | >19.8 |
| 3 | 4.4000 | >23.8 | >23.2 | >23.1 | >22.5 | >20.6 | >20.2 | >19.2 |
| 4 | 7.4756 | >25.4 | >25.1 | >24.3 | >23.9 | >21.8 | >21.2 | >19.8 |
| 5 | 11.4474 | >25.1 | >24.7 | >23.8 | >23.5 | >21.6 | >20.9 | >19.5 |
| 6 | 15.3913 | >23.5 | >23.5 | >22.9 | >22.7 | >20.8 | >20.5 | >19.6 |

**Note.** Column (2) provides the midtime of the observation, in units of days from burst trigger. Magnitudes are not corrected for Galactic foreground extinction. The photometry for epochs 1a–3 was obtained after image subtraction against epoch 4 (see Section 2.3 for details). On the contrary, the 3σ upper limits for epochs 4–6 have been measured directly, without image subtraction.

component was thus expected to be detectable with GROND within about 2 weeks after the GRB trigger. Therefore, following earlier efforts with GROND to explore GRB-SN light curves (E. F. Olivares et al. 2012, 2015; J. Greiner et al. 2015; D. A. Kann et al. 2019; S. Klose et al. 2019) we performed follow-up observations of the field during several epochs between 0.4 (A. Nicuesa Guelbenzu 2019) and 15.4 days post-GRB trigger (Table 1). Observations were then terminated since no evidence for SN light was found.

GROND data were reduced in a standard fashion (bias subtraction, flat-fielding, and coadding; T. Krühler et al. 2008; A. K. Yoldaş et al. 2008), based on numerical routines in the Image Reduction and Analysis Facility (IRAF; D. Tody 1993).

### 2.2. Large Binocular Telescope Late-time Observations

Late-time observations of the field were performed with the Large Binocular Telescope (LBT) using the two twin Large Binocular Camera instruments equipped with the Sloan filters g′r′i′z′ on 2023 October 20. These data were reduced using the data reduction pipeline developed at INAF–Osservatorio Astronomico di Roma (A. Fontana et al. 2014), which includes bias subtraction and flat-fielding, bad pixel and cosmic-ray masking, astrometric calibration, and coaddition.

On the LBT images, we measure the AB magnitudes for the host given in Table 4 (see Appendix A). Within the errors, these values are consistent with those measured with GROND in epoch 4 ($T - T_0 = 7.4756$ days). Based on these data we conclude that 7 days postburst the transient did not contribute anymore to the observed flux in g′r′i′z′ in a measurable quantity.

### 2.3. Image Subtraction and Photometry

In order to search for a transient hidden by the relatively bright host galaxy, we performed image subtraction on the GROND images using HOTPANTS (A. Becker 2015). HOTPANTS convolves the template and the source images to the same point-spread function (PSF) and photometric scale. As a template image we used the GROND fourth-epoch observations (Table 1) since these images have the best seeing (~1″.0). We aligned the template and the source images by means of wcsremap.[15]

The Gaussian input parameters in HOTPANTS were calculated following A. Becker (2015) considering the case in which the template FWHM is smaller than the source FWHM. In this way, the template images were smoothed to the seeing of the corresponding source images. In addition, we fixed several other parameters for HOTPANTS following L. Hu et al. (2022), but let free to vary the number of each region's stamps, the convolution kernel half width, and the size of Gaussians which compose the kernel. In doing so, we selected those input parameters that minimized the background noise on the residual image in a star-free region close to the position of the host galaxy. All resulting residual images were then photometrically calibrated with respect to the source images (see below). In the case of a nondetection of the transient the upper limit we used was measured locally on the subtracted image (Table 1). We have confirmed our results using also the fifth GROND epoch as a template (seeing ~1″.3), although we obtained a lower signal-to-noise ratio (SNR).

We double checked the detections using the late LBT images (about 4 yr or 1460 days after the burst trigger, see Section 2.2 and Table 4 in Appendix A). The g′i′z′ detections are confirmed for epochs 1a and 2, but with lower SNR likely due to the additional plate scaling, and a general worse PSF shape due to not perfect collimation. We have not checked the r′ band since the abovementioned problem was worse in this filter, although it did not compromise the aperture photometry in Table 4. For this filter we have instead used (as template) a stack of the best Gemini-S images[16] obtained at 58–63 days after the burst (see A. J. Levan et al. 2023), and we obtained consistent results. In other words, the results obtained with GROND could not be improved. Finally, to better constrain the late transient decay, we performed image subtraction on the Gemini r-band image at 11.4 days, using the later Gemini images as a template (see above), and obtained for the optical transient an upper limit of r > 25.8 AB mag.

Apparent magnitudes were obtained by performing aperture photometry using DAOPHOT and APPHOT under PyRAF/IRAF with increasing apertures up to 4 times the FWHM and by PSF photometry with the DAOPHOT and ALLSTAR tasks of IRAF. PSF fitting was used to measure the magnitudes of the transient after image subtraction. In the optical bands the data were calibrated using Pan-STARRS (K. C. Chambers et al.

---

[15] http://tdc-www.harvard.edu/wcstools

[16] Gemini images have been re-reduced using the dedicated Gemini pipeline DRAGONS (K. Labrie et al. 2019).

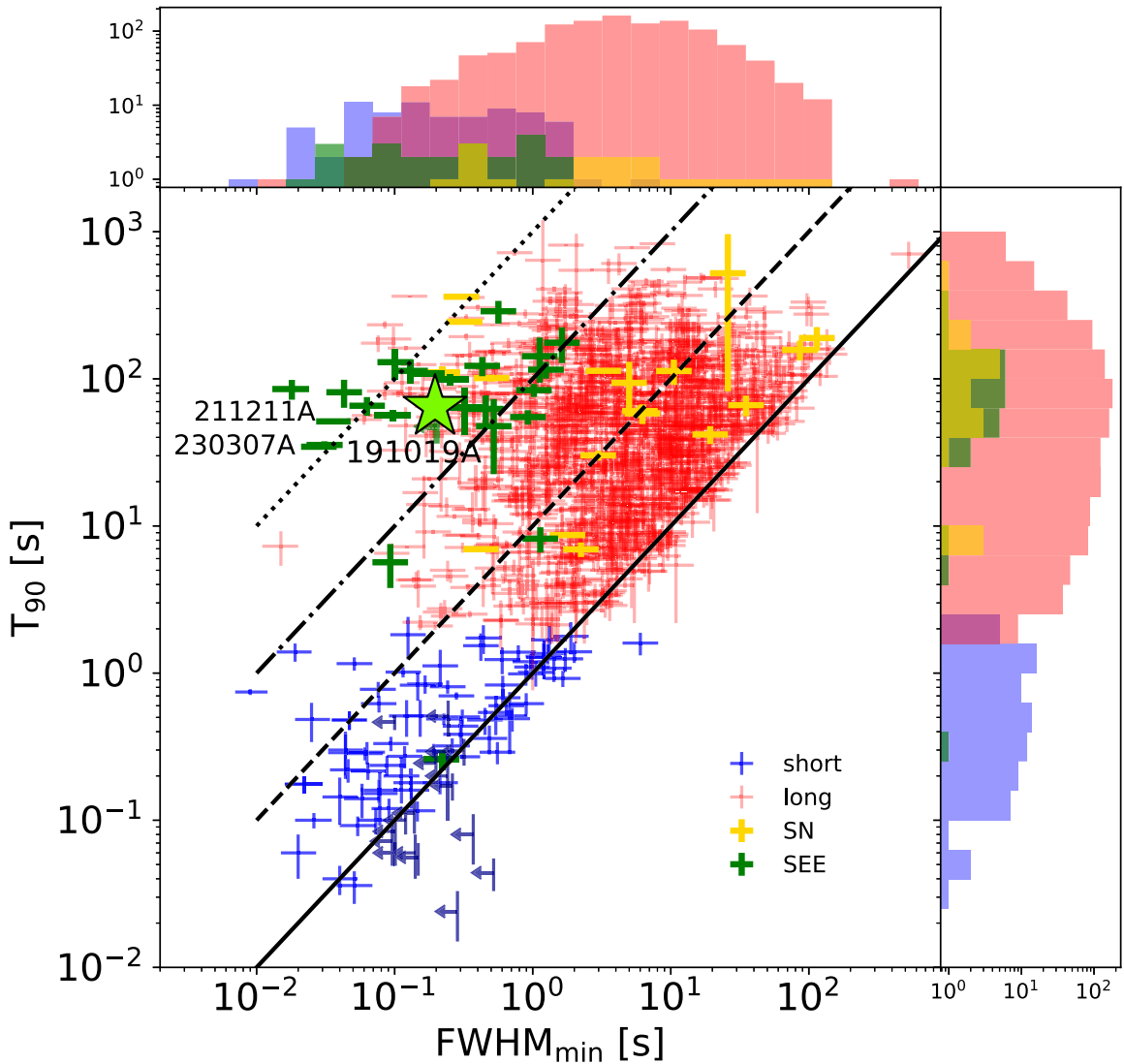

**Figure 1.** Burst duration vs. prompt emission minimum temporal variability of 1291 GRBs detected with Swift BAT (from 2005 January to 2022 July) and the corresponding marginal distributions. The sample includes 78 short GRBs (blue) and 24 short GRBs with SEE (green). Gold points are long GRBs with an associated SN (adapted from A. E. Camisasca et al. 2023). The location of GRB 191019A is indicated by a green star in the top-left side of the diagram where SEE GRBs are located, together with two long GRBs better identified with being originated from compact binary coalescence progenitors (GRB 211211A and GRB 230307A).

2016), in the near-infrared (NIR) bands using the Two Micron All-Sky Survey (M. F. Skrutskie et al. 2006). Central wavelengths of the GROND filter bands are listed in A. Rossi et al. (2011).

## 3. Results

### 3.1. Application of the Prompt Emission Minimum Variability Timescale Criterion

According to A. E. Camisasca et al. (2023), the GRB prompt emission MVT represents a diagnostic to infer the progenitor nature of a GRB. By estimating MVT from the FWHM of the shortest, statistically significant peak in the prompt emission light curve (FWHM$_{min}$) of hundreds of GRBs, it was found that high variability (low FWHM$_{min}$ values) typically belongs to compact binary mergers. Interestingly, this result was found to be independent of the burst duration and for this reason it is an important probe for the nature of peculiar long GRBs with properties more similar to those of short GRBs.

By analyzing the results from the Swift GRB sample analyzed by A. E. Camisasca et al. (2023), we checked which long GRBs sufficiently nearby for a possible kilonova detection, i.e., at $z < 0.5$, and without any associated SN, had an FWHM$_{min}$ compatible with compact binary merger values, and we found that GRB 191019A was satisfying our criteria. Specifically, for this GRB an MVT of FWHM$_{min} = 0.196^{+0.068}_{-0.05}$ s has been measured, which is compatible with the values typically found for GRBs associated with compact binary mergers (Figure 1). In particular, given the long burst duration of GRB 191019A ($T_{90} = 64.35 \pm 4.35$ s), its position in the FWHM$_{min}$ versus $T_{90}$ diagram lies among the short GRBs showing "soft extended emission" (SEE; e.g., J. P. Norris & J. T. Bonnell 2006; P. Y. Minaev et al. 2010). Interestingly, two famous long GRBs for which a kilonova component was found in the optical afterglow, namely GRB 211211A (J. C. Rastinejad et al. 2022; E. Troja et al. 2022; J. Yang et al. 2022; B. P. Gompertz et al. 2023) and GRB 230307A (e.g., A. J. Levan et al. 2023), lie in the same region of the diagram as GRB 191019A (Figure 1).

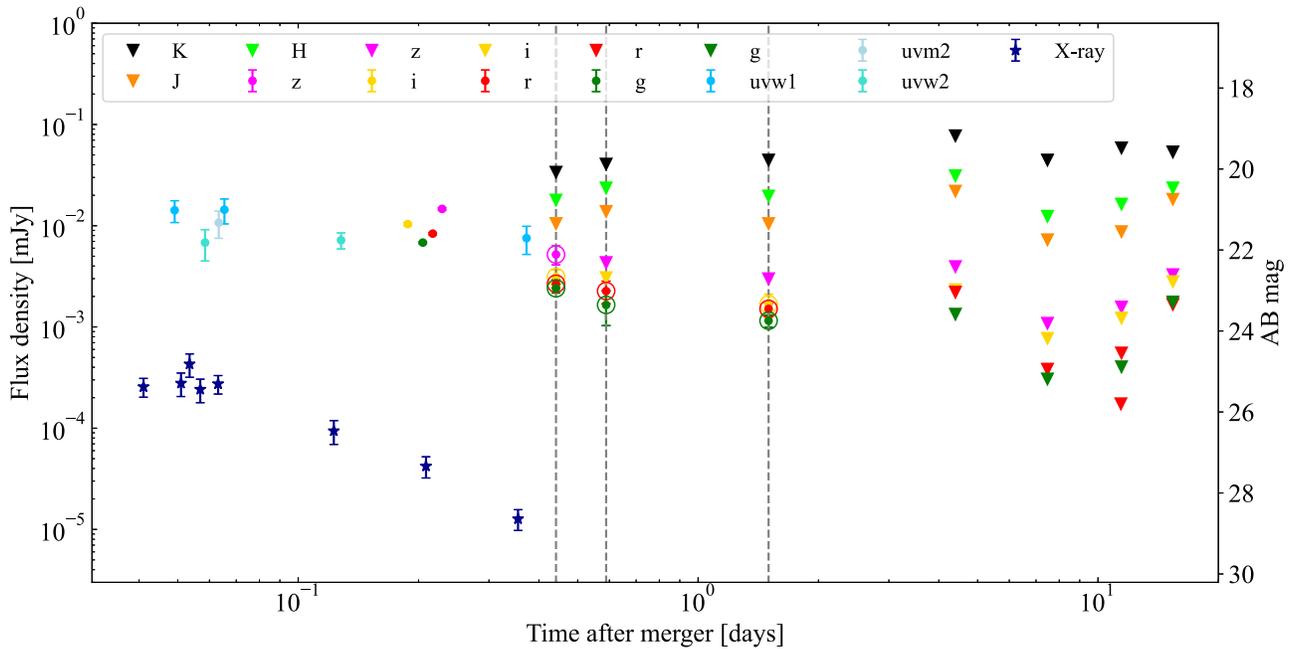

**Figure 2.** Multiband light curve of the transient that followed GRB 191019A. Optical and NIR data are from A. J. Levan et al. (2023, from their Table 4 of the supplementary material, where flux densities corrected for dust extinction and host-galaxy subtraction are quoted; we consider here only those with relative error ⩽ 30%) and from this work (larger markers) where GROND fluxes have been computed after image subtraction of the host galaxy observed at epoch 4 with GROND, and corrected for both Milky Way ($A_V = 0.10$ mag) and host-galaxy dust extinction ($A_V = 0.06$ mag). The vertical dashed lines mark the first three observation epochs quoted in Table 1. X-ray data were taken from the Swift-XRT GRB light-curve repository (P. A. Evans et al. 2007, 2009) as unabsorbed 0.3–10 keV fluxes computed with the Burst Analyzer tool (see P. A. Evans et al. 2010), and converted to 1 keV flux densities by assuming a power-law spectrum and the corresponding photon index provided in each temporal bin of the light curve.

The SEE distribution in the $T_{90}$–FWHM$_{min}$ plane clearly overlaps with the distribution of long GRBs associated with SNe, but at the same time the centroids of the two distributions are well separated: interestingly, GRB 191019A lies in the middle of the SEE region and in the outskirts of the SN-associated cases. This property alone cannot be taken as compelling evidence that GRB 191019A behaves like an SEE GRB, but suggests a probable SEE-like nature than a collapsar one. These findings further support past interpretation that also GRB 191019A originated from a compact binary merger (e.g., A. J. Levan et al. 2023), and address the possible presence of a kilonova component, similarly to GRB 211211A and GRB 230307A.

### 3.2. Reanalysis of the Light Curve of the Transient That Followed the Burst

#### 3.2.1. X-Rays

Adopting a power-law SED, the best fit of the Swift-XRT 0.3–10.0 keV spectrum, computed by the automatic algorithm of the UK Swift Science Data Center (UKSSDC) Swift-XRT GRB Spectrum Repository in the 3.2–32.4 ks temporal window,[17] provides a photon index of $\Gamma = 2.0^{+0.4}_{-0.3}$ and an intrinsic equivalent hydrogen column density of $N_{H,z} = 1.2^{+1.6}_{-1.2} \times 10^{21}$ cm$^{-2}$ in addition to a Galactic $N_H = 3.28 \times 10^{20}$ cm$^{-2}$. We obtained from the UKSSDC Burst Analyzer tool (P. A. Evans et al. 2010), the absorption-corrected X-ray fluxes in the energy range 0.3–10.0 keV, and we converted them into 1 keV flux densities by assuming a power-law spectrum and the corresponding photon index provided by the Burst Analyzer tool in each temporal bin of the light curve. The flux density evolution in the X-ray band is compatible with a power-law decay with $\alpha_X = 1.44 \pm 0.14$.[18]

#### 3.2.2. Optical Bands

We detected the optical transient during the first three GROND observations of the field, up to 1.5 days postburst (Table 1, Figure 2). The transient has coordinates R.A., decl. (J2000) = 22:40:05.861, −17:19:42.77 (±0″.3), measured on the combined $g'r'i'$ residual images of the second-epoch observations (Figure 3). Within the errors, these coordinates agree with what has been reported by D. A. Perley et al. (2019b).

Based on a joint fit of the griz-band magnitudes provided by A. J. Levan et al. (2023) and GROND's first visit of the field (epoch 1a in Table 1), we find that between both observing runs the optical transient was fading with a decay slope $\alpha_{opt} = 1.43 \pm 0.07$. Within the errors, $\alpha_X \sim \alpha_{opt}$ up to $T_0 + 0.4$ days (epoch 1a in Table 1), suggesting a common cooling regime of the electrons radiating optical and X-ray synchrotron light (see Figure 4). This finding is compatible with results by A. J. Levan et al. (2023), who showed that a single power-law SED from X-rays to the optical bands can describe the observations at $T_0 + 0.21$ days.

However, GROND's following observing runs 1b and 2 (Table 1) clearly show a significant flattening of the light curve, with a new decay slope of $\alpha_{opt} \sim 0.5 \pm 0.1$ (Figure 4). The rescaled X-ray power-law model underpredicts the flux in the optical bands, with a >3σ deviation observed at $T_0 + 1.5$ days. Even by assuming a shallower optical decay index (i.e., $\alpha_{opt} = \alpha_X - 0.25$), as for the case where the synchrotron cooling frequency lies between the X-ray and the optical regime (e.g., R. Sari et al. 1998), the discrepancy at late times (>1 day) persists (Figure 4).

---

[17] https://www.swift.ac.uk/xrt_spectra/algorithm.php

[18] For the time- and frequency-dependent flux density we use the notation $F(t, \nu) \propto t^{-\alpha}\nu^{-\beta}$.

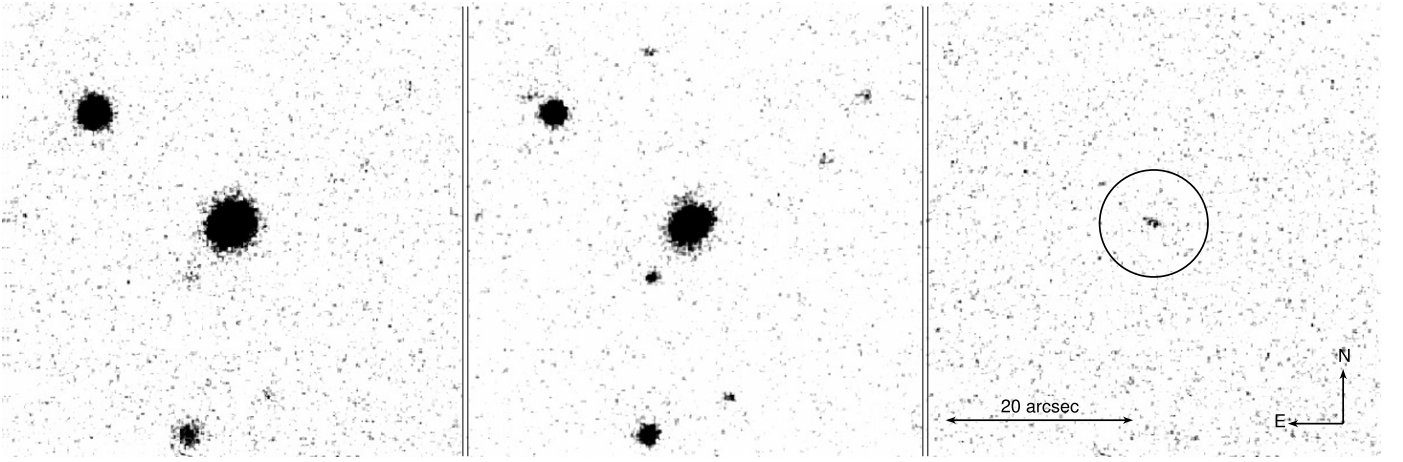

**Figure 3.** From left to right, GROND $r'$-band observations obtained during epochs 2 at 1.5 days, epoch 4 at 7.5 days, and the residual of the image subtraction (epoch 2 minus epoch 4) using `HOTPANTS`. The optical transient is indicated by the circle.

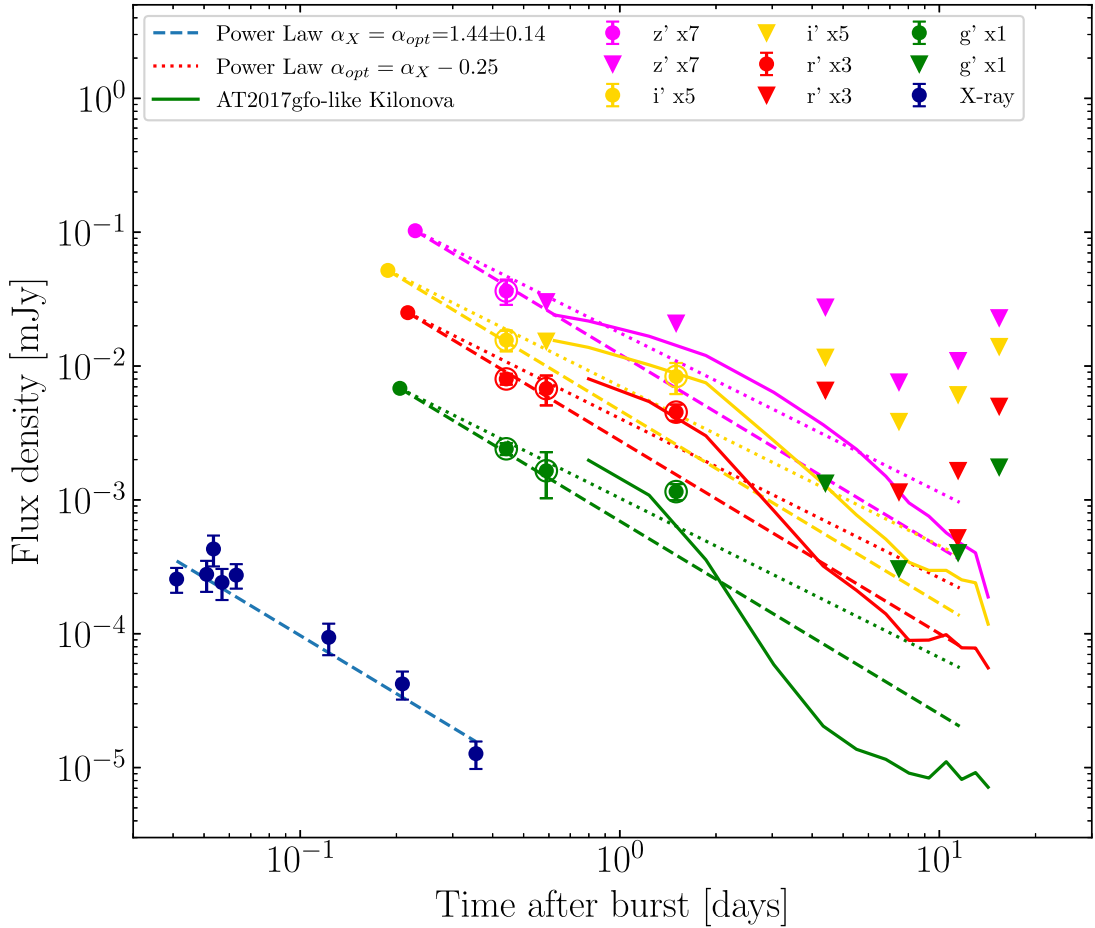

**Figure 4.** X-ray vs. optical light curve, where larger circles indicate GROND data from this work and the other data are taken from the literature (as in Figure 2). The blue dashed line shows the X-ray flux best-fit power-law model with a decay index $\alpha_X = 1.44 \pm 0.14$. This model is then scaled to match the optical data (dashed lines) together with a shallower power-law decay with $\alpha_X - 0.25$ (dotted lines). The two power-law models are expected if the optical emission is produced by electrons in the same cooling regime as those emitting in X-rays (dashed line) or in a different cooling regime (dotted line). The solid lines indicate the flux of a AT2017gfo-like kilonova if it were at the redshift of GRB 191019A and ~4 times more luminous. For visual purposes only, the $r'$-, $i'$-, and $z'$-band data have been rescaled by a factor of 3, 5, and 7, respectively.

To quantify the null hypothesis that a power-law model cannot fit the data and test for chromaticity of flux evolution, we considered the early data by A. J. Levan et al. (2023) and our GROND detections (i.e., from ~0.2 to 1.5 days) for the two filters with best temporal sampling (i.e., $r'$ and $g'$): a simple power-law model does not provide an acceptable fit, with probability of having the observed data set, given the null hypothesis ($p$-values), of $1.7 \times 10^{-5}$ and $4 \times 10^{-4}$, respectively. By assuming a broken power law, with a steep-to-shallow behavior and a temporal break at 0.44 days, we find a significant improvement of the fit, with $p$-values of 0.79 and 0.73 in the $r'$ and $g'$ bands, respectively. The best-fit broken power laws show that the flux evolution after the break is slightly shallower in the redder filter, suggesting achromatic behavior.

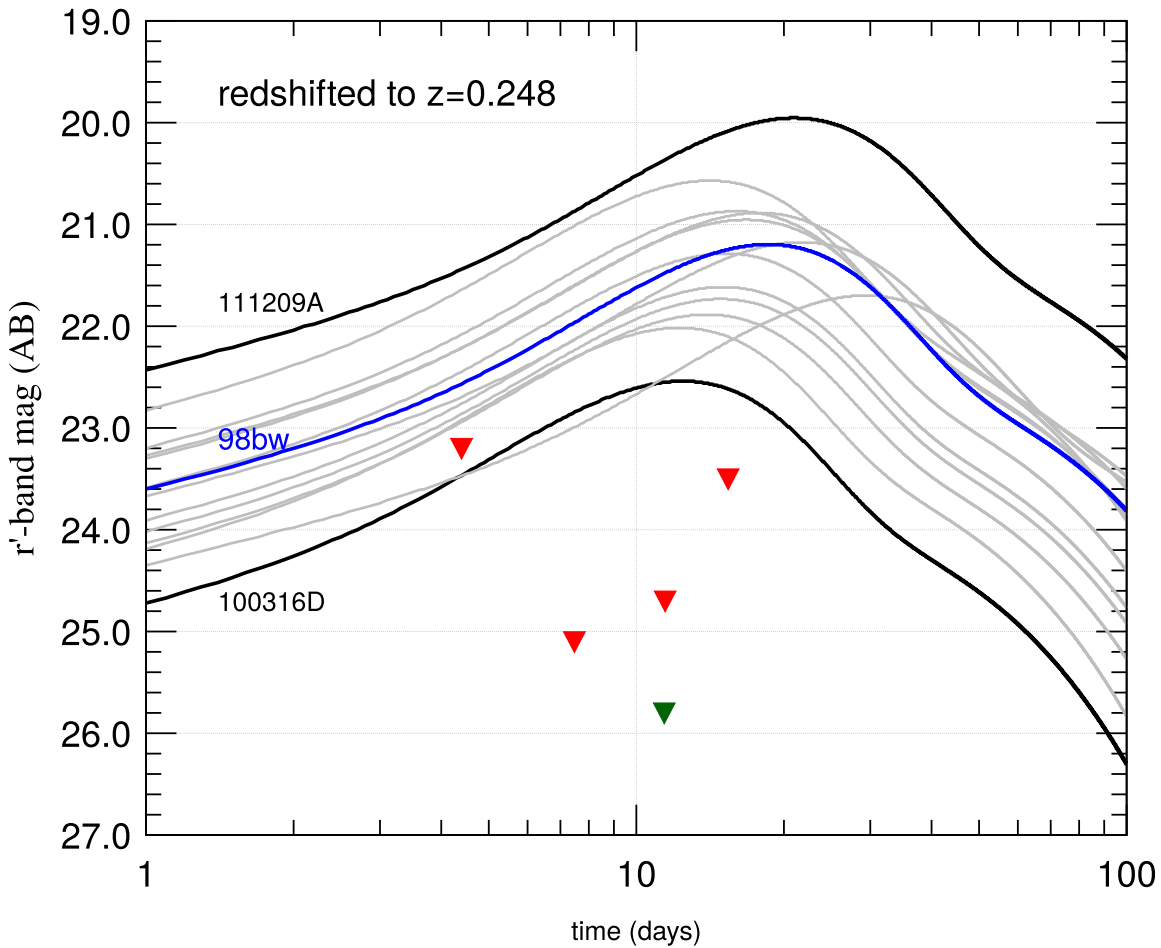

**Figure 5.** The $r'$-band light curves of all 13 GRB-SNe observed by GROND between 2007 and 2014, which are listed in Table 2 in S. Klose et al. (2019). All light curves have been calculated based on the luminosity and stretch factors found for these SNe and have been corrected for Galactic and host-galaxy extinction along the line of sight. They have been redshifted to $z = 0.248$ taking into account the appropriate cosmological corrections following the procedure described in A. Zeh et al. (2004). Also shown is the $r'$-band light curve of SN 1998bw (in blue) shifted to $z = 0.248$ taking into account the appropriate cosmological corrections. Overplotted are the upper limits we can set based on the GROND data (Table 1; $r'$ band) and the late Gemini observations (Section 2.3; dark green).

We note that this peculiar steep-to-shallow optical light-curve morphology is not unique, as it was already observed in a small subset of GRBs in the early years of the Swift era (see, e.g., D. A. Kann et al. 2010), although with light curves break on average much earlier than what we observe for GRB 191019A. For instance, the optical light curve of GRB 090102 (Gendre et al. 2010) is similar to GRB 191019A but the break time was ∼0.1 hr (rest frame) after the burst.

In the standard fireball model, which fairly accounts for late afterglow phenomenology as forward-shock radiation, a steep-to-shallow light curve is not envisioned (e.g., R. Sari et al. 1998). Such morphology could be produced by the presence of a reverse shock component dominating the forward shock at early times. However, a reverse shock is expected on timescales of the order of minutes (e.g., E. Nakar & T. Piran 2004), in contrast with what we observe for GRB 191019A. Alternatively, energy injection into the forward shock could cause a light-curve flattening. The spin-down radiation from a newly born magnetar (e.g., B. Zhang & P. Mészáros 2001) or a slightly off-axis structured jet (e.g., P. Beniamini et al. 2020) are plausible energy sources that are often invoked to explain the X-ray afterglow "plateaus" observed in GRBs, and which might have an optical counterpart. In the case of GRB 191019A the lack of any evidence of a shallow phase in X-rays cast some doubts against the energy injection scenario, however.

### 3.2.3. Constraints on Supernova Light

If GRB 191019A had been a classical long burst, then for a redshift of $z = 0.248$ an SN component was expected to be detectable with GROND. Observational constraints on SN light were first reported by A. J. Levan et al. (2023) based on HST observations 30 and 184 days post-GRB trigger. No evidence for transient emission was found ($g > 24$, $r > 23.5$, and $z > 22$ mag).

Figure 5 shows the upper limits we can set on any SN that followed GRB 191019A in comparison to the $r'$-band light curves of 13 GRB-SNe that have been observed with GROND

Table 2
The 90% Confidence Interval and Median Values of Each Parameter Inferred from the Simultaneous Afterglow (G. Ryan et al. 2020) and Kilonova (T. Dietrich et al. 2020) Modeling Performed by Using the NMMA Code

| $\log(E_0)$ (erg) (1) | $\log(n_0)$ (cm$^{-3}$) (2) | $\theta_c$ (rad) (3) | $\theta_w$ (rad) (4) | $\iota$ (rad) (5) | $p$ (6) | $\log(\epsilon_e)$ (7) | $\log(\epsilon_B)$ (8) | $\log(M_{\rm ej}^{\rm dyn})$ ($M_\odot$) (9) | $\log(M_{\rm ej}^{\rm wind})$ ($M_\odot$) (10) | $\Phi$ (deg) (11) |
|---|---|---|---|---|---|---|---|---|---|---|
| $52.14^{+0.37}_{-0.53}$ | $-0.14^{+1.38}_{-1.94}$ | $0.16^{+0.06}_{-0.05}$ | $0.38^{+0.19}_{-0.13}$ | $0.07^{+0.05}_{-0.03}$ | $2.75^{+0.04}_{-0.05}$ | $-0.69^{+0.39}_{-0.34}$ | $-4.60^{+1.56}_{-1.19}$ | $-1.73^{+0.25}_{-0.24}$ | $-1.36^{+0.39}_{-0.31}$ | $40.13^{+11.88}_{-10.30}$ |
| U (49, 53) | U (−3, 7) | U (0.01, π/10) | U (0.01, π/4) | Sine (0.0, π/8) | U (2.01, 3.0) | U (−5, 0) | U (−10, 0) | U (−3, −1) | U (−3, −0.5) | U (15, 75) |

**Note.** For the afterglow component, we assumed a Gaussian jet profile and fiducial values for the microphysical parameters (see Section 3.3 and Figures 6 and 7). The bottom part of the table quotes the assumed priors in the Bayesian inference analysis, where "U" stands for uniform function, with minimum and maximum values quoted in the brackets. $E_0$ = kinetic fireball energy; $n$ = particle number density in the circumburst environment; $\theta_c$ = half-opening angle of the jet core; $\theta_w$ = half-opening angle of the jet truncated-wings; $\iota$ = viewing angle with respect to jet axis; $p$ = electron energy distribution power-law index; $\epsilon_e$ = shock energy fraction that goes into the electrons; $\epsilon_B$ = shock energy fraction that goes into the magnetic energy density; $M_{\rm ej}^{\rm dyn}$ = dynamical ejecta mass; $M_{\rm ej}^{\rm wind}$ = wind ejecta mass; and $\Phi$ = half-opening angle of lanthanide-rich equatorial ejecta.

between 2007 and 2014 (see Table 2 in S. Klose et al. 2019). These events cover the whole GRB-SN luminosity distribution of well sampled GRB-SNe, from GRB 100316D/SN 2010bh (E. F. Olivares et al. 2012), one of the faintest GRB-SN ever detected (e.g., A. Melandri et al. 2014; Z. Cano et al. 2017; M. G. Dainotti et al. 2022), to GRB 111209A/SN 2011kl (D. A. Kann et al. 2019), the most luminous GRB-SN ever observed.

A visual inspection of Figure 5 shows that all light curves fall into a strip with a width of about 2–3 mag, which is limited by the very faint GRB 100316D/SN 2010bh and the very bright GRB 111209/SN 2011kl. The strongest limit we can set stamps from the time span between 7 and 12 days postburst and is well below the identified GRB-SN magnitude strip: if a CC-SN followed GRB 191019A, in $r'$ it was at least 3 mag less luminous than SN 1998bw and 2 mag fainter than SN 2010bh. According to the archived Gemini data (Section 2.3), the constraint on the luminosity is even stronger, >4 mag in $r$ at 11.4 days postburst.

In conclusion, even if we take into account that the time evolution of GRB-SN light curves shows a certain parameter range (characterized by a stretch factor), the SN limit we can set provides a constraint on the entire SN light curves. In this respect, any SN related to GRB 191019A must have been less luminous than each GRB-SN (see Appendix B).

### 3.3. Joint Afterglow and Kilonova Modeling

Since the flattening observed in the optical light curve at late times (epochs 1b and 2 in Table 1) cannot be explained by the presence of a slowly rising SN (Section 3.2.3), and probably not by a reverse shock or energy injection either (Section 3.2.2), we explore now the possibility that the source of this additional radiation component was kilonova light.

Given the relatively small redshift of GRB 191019A ($z = 0.248$), a kilonova with a luminosity similar to AT2017gfo is expected to lie within the discovery space of a 2 m telescope, if not hidden by an intrinsic low luminosity. Therefore, following the procedure described in A. Rossi et al. (2020), we compared our data at $T_0 + 1.5$ days in each filter, with the flux of the kilonova AT2017gfo (D. A. Coulter et al. 2017) associated with the NS–NS merger GW170817 (B. P. Abbott 2017), shifted to the redshift of GRB 191019A. We find that an emission component similar to AT2017gfo but ~4 times more luminous could reproduce the observations in the $i'$ and $r'$ bands, while our $g'$-band flux seems to be brighter (Figure 4). Apparently, the presence of a kilonova associated with GRB 191019A is overall compatible with the general properties of the observed optical–NIR emission at 0.4 and 1.5 days, although with slightly different brightnesses than AT2017gfo.[19]

Motivated by the results obtained with our previous simplistic approach, we then explored the possible presence of a kilonova by comparing our multiband data set ($g'$, $r'$, $i'$, $z'$, and X-ray) with a much more sophisticated model that takes into account simultaneously the presence of an afterglow and a kilonova component through a joint fit. For this purpose, we exploited the NMMA v0.2.0 framework, which allows us to estimate best-fit parameters with a Bayesian inference method (T. Dietrich et al. 2020; P. T. H. Pang et al. 2023).

For the afterglow modeling, NMMA uses the Afterglowpy Python module (G. Ryan et al. 2020). Afterglowpy allows one to model GRB afterglow light curves and spectra by taking into account the possible effects due to a complex jet structure and an off-axis observer. We here assumed a Gaussian profile for the jet structure, and that the whole ($\xi_N = 1$) electron population is shock accelerated. With the exception of the viewing angle $\iota$, which has a sine function, to all parameters we assigned a uniform function to model the prior probability (for details see Table 2).

For the kilonova emission, NMMA allows one to fit and simulate data using several numerical and analytical models. Assuming an NS–NS merger as the progenitor system, we adopted here the kilonova modeling resulting from the time dependent 3D Monte Carlo code POSSIS for modeling radiation transport (M. Bulla 2019, 2023). We used the kilonova model grid published in T. Dietrich et al. (2020) based on the first version of POSSIS (M. Bulla 2019) where the kilonova ejecta are represented with two components: (1) a high-velocity ($0.08 < v_{\rm dyn}/c < 0.3$) dynamical ejecta of mass $M_{\rm ej}^{\rm dyn}$ with a lanthanide-rich composition distributed about the equatorial plane with half-opening angle $\Phi$, and lanthanide-poor composition at higher latitudes, and (2) a slower ($0.025 < v_{\rm wind}/c < 0.08$) wind (or "postmerger") component, which is spherical ejecta released from the merger remnant and

---
[19] Upper limits computed at late epochs are too shallow to provide any useful constraint to the comparison with AT2017gfo.

Table 3
Expected and Observed Supernova AB Magnitudes

| $dt$ (days) | $r'$ exp (mag) | $r'$ obs (mag) | diff | $i'$ exp (mag) | $i'$ obs (mag) | diff | $z'$ exp (mag) | $z'$ obs (mag) | diff |
|---|---|---|---|---|---|---|---|---|---|
| (1) | (2) | (3) | (4) | (5) | (6) | (7) | (8) | (9) | (10) |
| 7.48 | 22.0 | 25.1 | 3.1 | 22.1 | 24.3 | 2.2 | 22.2 | 23.9 | 1.7 |
| 11.45 | 21.5 | 24.7 | 3.2 | 21.6 | 23.8 | 2.2 | 21.8 | 23.5 | 1.8 |
| 15.39 | 21.3 | 23.5 | 2.2 | 21.3 | 22.9 | 1.6 | 21.5 | 22.7 | 1.2 |

**Note.** All observed ("obs") AB magnitudes in the $r'$, $i'$, and $z'$ filters are $3\sigma$ upper limits (Figure 5). Columns (2), (5), and (8) quote the expected ("exp") magnitudes. All the corresponding differences (observed minus expected; columns (4), (7), and (10)) are lower limits.

debris disk, with an intermediate lanthanide content and mass $M_{\rm ej}^{\rm wind}$ (see T. Dietrich et al. 2020). In the fit we fixed the GRB 191019A luminosity distance at 1289.3 Mpc, as obtained from the measured redshift ($z = 0.248$) and assuming a flat cosmological model (see Section 1). We also considered an additional error budget (in magnitudes), which takes into account the uncertainties on the model predictions as well as possible systematic errors ($em_{\rm syserr}$ in the corner plot), to which we assigned a uniform prior with range from 0 to 2 mag. The resulting best-fit light curves computed in each band are plotted in Figure 6, the fit corner plot is in Figure 7, while the 90% confidence interval and median values of each parameter, as well as the adopted prior functions, are quoted in Table 2.

By simply inspecting the resulting light curves, it is evident that the afterglow component is constrained mostly by the X-ray data and it is dominant during the early epochs in the optical bands. The afterglow best fit jet core semiaperture angle is about ~10° ($9^{+4}_{-3}$ deg), and the observer viewing angle is ~4° ($4^{+3}_{-2}$ deg), thus within the jet core. The fireball isotropic kinetic energy $E_0$ ($1.4^{+1.8}_{-1.0} \times 10^{52}$ erg) is nicely compatible with the observed prompt emission equivalent isotropic energy $E_{\rm iso}$, by assuming a reasonable efficiency of about $\eta \sim 10\%$, where $\eta = \frac{E_{\rm iso}}{E_{\rm iso} + E_0}$. Indeed, the 15–150 keV fluence measured by Swift BAT is $(1.00 \pm 0.03) \times 10^{-5}$ erg cm$^{-2}$ (H. A. Krimm et al. 2019), corresponding to a radiated energy $E_{\rm iso} = (1.70 \pm 0.05) \times 10^{51}$ erg. The latter was computed by assuming a power-law spectrum in the BAT bandpass: this assumption is reasonable given the softness of the spectrum, which is best fit with a photon index $\Gamma = 2.25 \pm 0.05$ (H. A. Krimm et al. 2019), suggesting that the prompt emission peak energy lies below the BAT bandpass.

Figure 6 also clearly shows that at later epochs (>1 day), the observed flux lies above the predicted levels of the afterglow component, and the presence of a kilonova provides a better match. Indeed, by assuming only the afterglow model, i.e., by removing the kilonova component from our initial model, we obtained a worse fit, with Bayesian evidence $\ln(Z) = -21.1$. By considering the ratio with the Bayesian evidences of the joint afterglow and kilonova model, for which we obtain $\ln(Z_0) = -14.1$, the resulting Bayes factor ($\ln(B) = \ln(Z/Z_0) = -7.0$) indicates a strong preference for the model which includes the kilonova component (see, e.g., N. Kunert et al. 2024 and references therein).

The kilonova properties we find from our fit are compatible with a dynamical ejecta mass of $M_{\rm ej}^{\rm dyn} \sim 0.02\ M_\odot$ and a wind mass $M_{\rm ej}^{\rm wind} \sim 0.04\ M_\odot$, though with large uncertainties (relative errors $\geqslant 60\%$). These values are slightly higher (yet consistent within the uncertainties) with those found for AT2017gfo by assuming the same kilonova modeling (T. Dietrich et al. 2020). This is in line with the need for a brighter kilonova (of about a factor 4; see Section 3.3 and Figure 4) by simply superposing an AT2017gfo-like light curve on the data.

We stress here that the real picture is likely much more complicated than a two-component scenario, and the dynamical/wind masses inferred should rather be interpreted as belonging to some high-/low-velocity components of a multicomponent scenario with multiple ejecta episodes (e.g., S. Bernuzzi 2020; V. Nedora et al. 2021). At the same time, the kilonova model we assumed is among the most sophisticated ones publicly available, and the assumption of a kilonova from an NS–NS merger progenitor is the most reasonable choice given that, so far, the only GRB with kilonova emission for which we were able to infer the progenitor nature was GRB 170817/AT2017gfo, which we know from gravitational-wave data analysis as originating from an NS–NS merger. Nevertheless, we explored also other kilonova models available within the NMMA framework, with different levels of sophistication, and which have different assumptions on the progenitor nature and on the kilonova ejecta properties (see Appendix C). We find that the data set for GRB 191019A did not allow us to confidently distinguish among different kilonova models, apart disfavoring the most simplistic ones. However, in all cases, we find that a joint afterglow plus kilonova fit is preferred with respect to an afterglow-only model.

Interestingly, both the afterglow-only and afterglow plus kilonova model provide as a best fit for the circumburst environment particle density a value $n_0 < 1$ cm$^{-3}$, which is typically deduced for short-GRB environments (e.g., E. Berger 2014; W. Fong et al. 2015). This result is in stark contrast with the interpretation by D. Lazzati et al. (2023), which invokes the presence of a very high-density environment with $n_0 \sim 10^7$–$10^8$ cm$^{-3}$ (see also A. J. Levan et al. 2023).

## 4. Discussion and Conclusions

Recent analysis of an increasing number of GRBs which were initially classified as long bursts, and therefore thought to be originating from collapsing massive stars, suggest a better compatibility of the observational data with compact binary merger progenitors. By ignoring the standard burst duration classification, typical features of GRBs associated with compact binary mergers are (1) the absence of an associated CC-SN, (2) the presence of an optical–NIR rebrightening compatible with a kilonova, (3) an early-type host galaxy, and (4) a distant GRB explosion site with respect to the corresponding galaxy center. GRB 191019A was suggested to belong to this sample of misclassified long GRBs, given the nondetection of any SN component (A. J. Levan et al. 2023) despite its close cosmological distance ($z = 0.248$), and the evidence of a host galaxy dominated by an old stellar population. However, the proximity of the GRB 191019A explosion site to the photometric center of its host galaxy ($\lesssim 100$ pc) is at odds with typical short-GRB galactocentric distances, and could suggest a compact binary formed in the dense circumnuclear disk of an active galactic nucleus (AGN). In this scenario (D. Lazzati et al. 2022) a very high-density environment ($n_0 > 10^7$ cm$^{-3}$) could be the origin of the long duration of the prompt emission (see D. Lazzati et al. 2023). Given the low distance of this burst, kilonova light is expected to be seen. However, the temporal coverage and limiting magnitudes of previous data sets did not allow us to address this question yet.

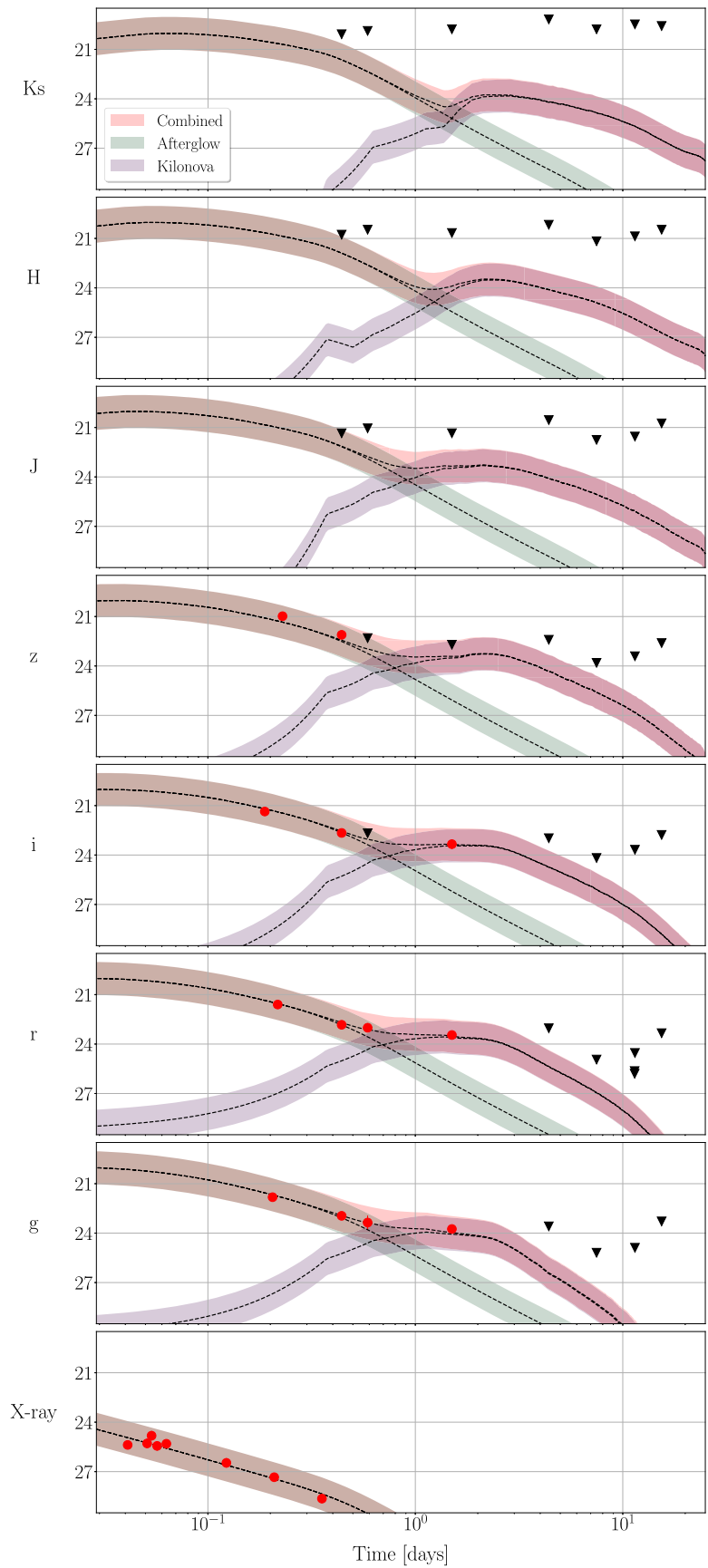

**Figure 6.** Joint fit of an afterglow plus a kilonova model, performed with NMMA (see Table 2). The dashed lines show the median light curve, while the shaded areas show the 95% interval. Red circles and black triangles mark the detections and upper limits, respectively, in AB magnitudes.

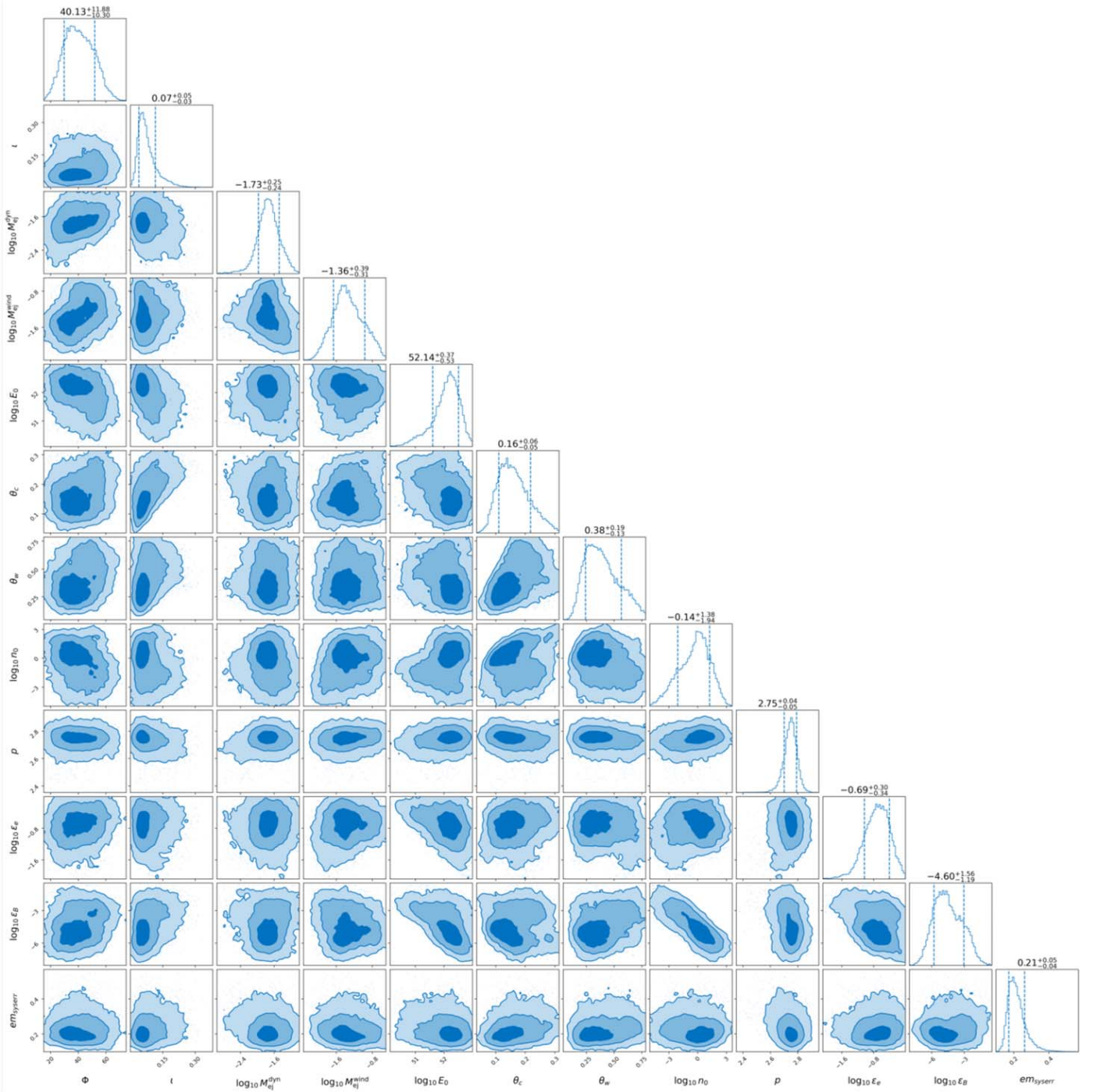

**Figure 7.** Corner plot obtained with NMMA from a joint fit assuming a model that incorporates both an afterglow and a kilonova component (see Table 2 and Figure 6). Different shadings mark the 68%, 95%, and 99% confidence intervals. For the 1D posterior probability distributions, the 90% confidence interval (dashed lines) and median values above each panel are indicated.

In this work, we further investigate the burst properties of GRB 1901910A and the nature of the optical transient that followed the burst, with the goal to find arguments in favor or against its possible compact merger origin. We found independent support of the compact binary merger progenitor nature from the high-energy prompt emission temporal variability, which has been recently found to be an additional potential diagnostic to infer the progenitor nature, where high variability values suggest a compact binary merger progenitor (A. E. Camisasca et al. 2023). We noticed that, for GRB 191019A, the obtained MVT value of ~20 ms locates this burst close to the centroid of the distribution of those GRBs associated with compact binary merger progenitors (A. E. Camisasca et al. 2023 and Figure 1). The same variability properties were found also for two other long GRBs with evidence of kilonova (GRB 111210A and GRB 230307A). These results independently support past interpretations on the progenitor nature of GRB 191019A and address to the possible presence of a kilonova component in its optical transient.

An optical transient following GRB 191019A was at first reported by D. A. Perley et al. (2019b) but its nature could not be clearly pinned down. Using GROND multicolor data obtained between 0.4 and 15 days postburst, we argued here that the temporal evolution of the transient's brightness disfavors pure afterglow emission (Figure 4). While the GROND data confirm the absence of an SN component (Figure 5), we found that the luminosity evolution of the optical transient is in agreement with an afterglow plus a kilonova signal compatible with AT2017gfo redshifted to the distance of the GRB host galaxy, though slightly brighter by a factor of a few.

By modeling the afterglow and the kilonova components, we were able to estimate the kilonova ejecta mass, which is slightly higher but still consistent, within the large uncertainties, with the one measured for AT2017gfo with similar assumptions on the progenitor system (an NS–NS) and on the kilonova modeling (e.g., T. Dietrich et al. 2020). By assuming other kilonova models with different levels of complexity and different progenitor assumptions, the data did not allow us to discriminate among them, apart disfavoring the most simplistic one. However, in all cases, results strongly favor the presence of a kilonova with respect to an afterglow-only model (Table 5).

Our findings strongly suggest that GRB 191019A might belong to the increasing list of long GRBs with an associated kilonova, beside GRB 211211A and GRB 230307A, and other cases with less robust evidence, as for instance GRB 060614 (e.g., Z.-P. Jin et al. 2015). Another interesting finding from our analysis, is the evidence of a low circumburst density (Table 2). This result is in stark contrast with the one obtained by D. Lazzati et al. (2023) from prompt emission light-curve modeling, where extreme values of $n_0 > 10^7$ cm$^{-3}$ were found. These high-density values were considered consistent with the apparent position of the GRB, very close to the center of an AGN-like host galaxy (though no direct proof on the existence of an AGN have been provided yet; see A. J. Levan et al. 2023). The interaction between a jet and an AGN dense circumnuclear environment, however, may choke or strongly suppress GRB relativistic jets, unless particularly bright and long, as also supported by recent studies (e.g., H.-H. Zhang et al. 2024). GRB 191019A is classified as a normal burst, with a well detected optical counterpart (see, e.g., H.-H. Zhang et al. 2024), and with low dust extinction along the line of sight within the host ($A_V^{host}$ in the range from 0.06 to 0.10 mag; A. J. Levan et al. 2023), further supported by UV-band detections (S. J. LaPorte et al. 2019). These properties are more consistent with a low-density environment, as the one we obtained from a joint fitting of an afterglow and a kilonova component model to the X-ray and optical data of GRB 191019A. More in general, if GRB 191019A is a disguised short GRB with a compact binary merger origin, as both past studies and this work suggest, a low-density environment is in agreement with the typical density values found in short GRBs and with the environments expected for compact binary mergers.

In the low-density scenario, the very small offset from the host galaxy of GRB 191019A, can be explained with a negligible transverse projection of a GRB located far off center. In this scenario, the long duration of the prompt emission, which D. Lazzati et al. (2023) explained as caused by the extremely dense circumburst environment, may have instead an intrinsic origin. An interesting hypothesis that was proposed in the past to explain short GRBs with SEE (e.g., M. V. Barkov & A. S. Pozanenko 2011; S. Kisaka & K. Ioka 2015) and which has been recently studied in much greater detail (e.g., C. Musolino et al. 2024), invokes fallback accretion onto the remnant compact object. Whether this is the case for GRB 191019A goes beyond the scope of this work and will be addressed in another study.

## Acknowledgments

We thank P. Pang for his precious support on NMMA and useful discussions. A.N.G. acknowledges logistic support by the Thüringer Landessternwarte Tautenburg, Germany. G.S. and P.S acknowledge the support by the State of Hesse within the Research Cluster ELEMENTS (Project ID 500/10.006). A. R. and E.P. acknowledge support by PRIN-MIUR 2017 (grant No. 20179ZF5KS). Part of the funding for GROND (both hardware and personnel) was generously granted by the Leibniz-Prize to G. Hasinger (DFG grant HA 1850/28-1) and by the Thüringer Landessternwarte Tautenburg. S.B. acknowledges funding from the EU Horizon under ERC Consolidator grant, no. InspiReM-101043372. A.E.C. is partially supported by the 2023/24 "Research and Education" grant from Fondazione CRT. The research leading to these results has received funding from the European Union's Horizon 2020 Programme under the AHEAD2020 project (grant agreement No. 871158).

The OAVdA is managed by the Fondazione Clément Fillietroz-ONLUS, which is supported by the Regional Government of the Aosta Valley, the Town Municipality of Nus, and the Unité des Communes valdotaines Mont-Emilius. The LBT is an international collaboration of the University of Arizona, Italy (Istituto Nazionale di Astrofisica, INAF), Germany (LBT Beteiligungsgesellschaft, LBTB), and The Ohio State University, representing also the University of Minnesota, the University of Virginia, and the University of Notre Dame. This work made use of data supplied by the UK Swift Science Data Centre at the University of Leicester.

*Facilities:* Max Planck:2.2m, Swift (BAT, XRT, and UVOT), and LBT.

*Software*: Afterglowpy (G. Ryan et al. 2020), NMMA (P. T. H. Pang et al. 2023), astropy (Astropy Collaboration et al. 2013, 2018), HOTPANTS (A. Becker 2015), PyRAF (Science Software Branch at STScI 2012), DRAGONS (K. Labrie et al. 2019), and WCSTools (J. Mink 2019).

## Appendix A
## Host Galaxy

Table 4 provides the magnitude of the GRB host galaxy based on GROND's fourth-epoch observations and LBT's visit of the field 4 yr after the burst (see Section 2.2).

Table 4
Measured AB Magnitudes of the Host

| Time (days) (6) | Filter (7) | Epoch (8) | Flux (mag(AB)) (9) | Telescope (10) |
|---|---|---|---|---|
| Host observed by GROND | | | | |
| 7.4756 | $g'$ | 4 | 20.04 ± 0.03 | GROND |
| 7.4756 | $r'$ | 4 | 19.03 ± 0.02 | GROND |
| 7.4756 | $i'$ | 4 | 18.65 ± 0.01 | GROND |
| 7.4756 | $z'$ | 4 | 18.44 ± 0.02 | GROND |
| Host observed by LBT | | | | |
| ~1460 | $g'$ | ... | 20.00 ± 0.01 | LBT |
| ~1460 | $r'$ | ... | 19.04 ± 0.01 | LBT |
| ~1460 | $i'$ | ... | 18.66 ± 0.01 | LBT |
| ~1460 | $z'$ | ... | 18.45 ± 0.02 | LBT |

**Note.** Magnitudes are measured within an aperture with a radius of 4 × FWHM and are not corrected for Galactic foreground extinction.

## Appendix B
## Gamma-Ray Burst Supernova Peak Luminosity Range

We further investigated on possible biases toward the faint end of our GRB-SN sample by comparing with Type I SN peak luminosity properties observed so far. The $r$-band absolute peak magnitude ($M_{r,\text{peak}}$) distribution of broad-line (BL) Type Ic SNe spans at most 2.5 mag (between −17.7 and −20.8 mag, with an average of $M_{r,\text{peak}} = -18.7 \pm 0.7$ mag; F. Taddia et al. 2019; C. Barbarino et al. 2021; S. Gomez et al. 2022). GRB-SNe observed to date with good data sampling are at least ~1 mag brighter and trace the high-luminosity end of the above Type Ic BL distribution (e.g., J. Hjorth & J. S. Bloom 2012; J. Hjorth 2013; Z. Cano et al. 2017; S. Klose et al. 2019), but the width of their peak luminosity distribution (J. Hjorth & J. S. Bloom 2012; Z. Cano et al. 2017; D. A. Kann et al. 2019; S. Klose et al. 2019) is not substantially different to the width of the corresponding peak luminosity distribution of Type Ic and Ic BL SNe (see also A. M. Soderberg et al. 2006). By considering GRB-SNe with the best follow-up in multiple bands (including spectroscopy), on the high end site of the peak luminosity distribution remains GRB 111209A/SN 2011kl (e.g., D. Nakauchi et al. 2013; D. A. Kann et al. 2019) while on the low-end side there are GRB 100316D/SN 2010bh (e.g., R. L. C. Starling et al. 2011; E. F. Olivares et al. 2012) and GRB 060218/SN 2006aj (e.g., P. Ferrero et al. 2006; E. Pian et al. 2006): the $r$-band peak magnitudes of these extreme GRB-SNe have a difference that lies between 2 and 3 mag, similar to Type Ic BL SNe. In principle, the long bursts GRB 990712 (G. Björnsson et al. 2001), GRB 021211 (M. Della Valle et al. 2003; A. Zeh et al. 2004), GRB 040924 (A. M. Soderberg et al. 2006; K. Wiersema et al. 2008), and GRB 060904B (Z. Cano et al. 2017) could have been followed by even slightly fainter SNe than SN 2010bh, but in these cases the database is comparably poor and no strong conclusions could be made. While, e.g., any SN that was associated with the long bursts GRB 060605 and GRB 060614 must have had a luminosity <1% of the luminosity of the prototypical SN 1998bw (J. P. U. Fynbo et al. 2006), here and in similar cases (e.g., GRB 111005A; M. J. Michałowski et al. 2018; M. Tanga et al. 2018) there is no evidence for SN light.

In conclusion, if substantially less luminous GRB-SNe do exist formally remains an open question, at least from the observational point of view, and the peak luminosity range of the GRB-SN sample we have used to infer the luminosity of any SN associated with GRB 191019A is the most robust so far.

## Appendix C
## Afterglow and Kilonova Joint-fit Comparison

In Table 5 we present the results obtained for different kilonova models within the NMMA framework, in addition to the kilonova model presented in Section 3.3.

At first, we have tested a model from D. Kasen et al. (2017, Kasen17-Jet in Table 5) which assumes only one ejecta component and has three parameters: the ejecta mass, the ejecta velocity, and the lanthanide mass fraction ($\chi_{lan}$). Then we made a different assumption on the progenitor by considering a kilonova emission from an NS–black hole binary system as predicted with POSSIS (Bu19-NSBH-Jet in Table 5), which has three parameters: the dynamical ejecta mass, the wind mass, and the orbital plane inclination, which is linked to the viewing angle of the jet assumed in the afterglow modeling. These models were finally compared with the results obtained by assuming the simple analytical model described in B. D. Metzger (2017; Metzger17-Jet in Table 5), which assumes one component and has four parameters: the ejecta mass, the ejecta velocity, the power-law index $\beta$ of the ejecta mass distribution expressed as a function of its velocity (the faster ejecta/matter lies ahead of slower matter and the distribution of mass with velocity greater than the value $v_0$ can be approximated with a power law $M(>v_0) = M(v/v_0)^{-\beta}$), and the opacity $k_r$.

The afterglow emission was modeled within the `Afterglowpy` framework (Afterglow in Table 5), as described in Section 3.3. However, contrary to our previous analysis, the microphyiscal parameters $\epsilon_e$ and $\epsilon_B$ are now fixed to 0.5 and 0.01, respectively. In doing so, we aim at reproduce similar assumptions to D. Lazzati et al. (2023) and further investigate on the circumburst density estimates.

The highest Bayesian evidence ($Z$) is obtained for the Bu19-NSBH-Jet model. Following N. Kunert et al. (2024) and references therein, by considering Bu19-NSBH-Jet as the reference model, we find $-1.10 < \ln(Z/Z_{\text{Bu19-NSBH-Jet}}) < 0$ for the Kasen17-Jet and Bu19-BNS-Jet models, while $\ln(Z/Z_{\text{Bu19-NSBH-Jet}}) < -4.5$ for the Metzger17-Jet and Afterglow models, indicating strong evidence against the latter two models, while no preference could be set among the Bu19-BNS-Jet, Bu19-NSBH-Jet, and Kasen17-Jet models.

Table 5
The 90% Confidence Interval and Median Values for Each Parameter Inferred from Joint Afterglow and Kilonova and Afterglow-only Jet Modeling Performed by Using the NMMA Code

| | | Bu19-BNS-Jet | Bu19-NSBH-Jet | Kasen17-Jet | Metzger17-Jet | Afterglow | Prior |
|---|---|---|---|---|---|---|---|
| $\log(E_0)$ | (erg) | $50.55^{+0.30}_{-0.21}$ | $50.58^{+0.25}_{-0.22}$ | $50.37^{+0.32}_{-0.13}$ | $50.91^{+0.35}_{-0.64}$ | $50.44^{+0.15}_{-0.14}$ | U (49, 53) |
| $\log(n_0)$ | (cm$^{-3}$) | $-2.14^{+0.70}_{-0.86}$ | $-1.99^{+0.86}_{-0.89}$ | $-1.49^{+0.39}_{-1.04}$ | $-3.01^{+1.74}_{-1.05}$ | $-0.27^{+0.21}_{-0.32}$ | U (−3, 7) |
| $\theta_c$ | (rad) | $0.21^{+0.05}_{-0.10}$ | $0.22^{+0.05}_{-0.10}$ | $0.25^{+0.04}_{-0.05}$ | $0.21^{+0.05}_{-0.04}$ | $0.23^{+0.04}_{-0.04}$ | U (0.01, π/10) |
| $\theta_w$ | (rad) | $0.39^{+0.20}_{-0.18}$ | $0.50^{+0.12}_{-0.26}$ | $0.53^{+0.14}_{-0.16}$ | $0.51^{+0.20}_{-0.18}$ | $0.57^{+0.15}_{-0.25}$ | U (0.01, π/4) |
| $\iota$ | (rad) | $0.05^{+0.13}_{-0.03}$ | $0.17^{+0.08}_{-0.14}$ | $0.12^{+0.06}_{-0.06}$ | $0.10^{+0.09}_{-0.05}$ | $0.26^{+0.05}_{-0.05}$ | Sine(0.0, π/8) |
| $p$ | | $2.72^{+0.06}_{-0.06}$ | $2.73^{+0.06}_{-0.11}$ | $2.71^{+0.06}_{-0.09}$ | $2.70^{+0.06}_{-0.07}$ | $2.23^{+0.16}_{-0.18}$ | U (2.01, 3.0) |
| $\log(M_{\rm ej}^{\rm dyn})$ | ($M_\odot$) | $-1.74^{+0.30}_{-0.33}$ | $-1.60^{+0.25}_{-0.22}$ | ... | ... | ... | U (−3, −1) |
| $\log(M_{\rm ej}^{\rm wind})$ | ($M_\odot$) | $-1.28^{+0.46}_{-0.47}$ | $-1.03^{+0.27}_{-0.39}$ | ... | ... | ... | U (−3, −0.5) |
| $\log(M_{\rm ej})$ | ($M_\odot$) | ... | ... | $-1.67^{+0.29}_{-0.35}$ | $-0.88^{+0.21}_{-0.61}$ | ... | U (−3, −0.5) |
| $\log(v_{\rm ej})$ | (c) | ... | ... | $-1.06^{+0.18}_{-0.21}$ | $-1.00^{+0.18}_{-0.30}$ | ... | U (−2, −0.5) |
| $\Phi$ | (deg) | $44.56^{+11.19}_{-13.04}$ | ... | ... | ... | ... | U (15, 75) |
| $\log(\epsilon_e)$ | | −0.3 | −0.3 | −0.3 | −0.3 | −0.3 | fixed |
| $\log(\epsilon_B)$ | | −2.0 | −2.0 | −2.0 | −2.0 | −2.0 | fixed |
| $\log(\chi_{\rm lan})$ | | ... | ... | $-5.86^{+0.72}_{-1.25}$ | ... | ... | U (−9, −1) |
| $\log(\kappa_r)$ | | ... | ... | ... | $-0.60^{+0.43}_{-0.22}$ | ... | U (−1, 2) |
| $\beta$ | | ... | ... | ... | $3.98^{+0.65}_{-1.53}$ | ... | U (1, 5) |
| $\ln(Z)$ | | −13.5 | −12.4 | −12.9 | −18.7 | −19.2 | ... |

**Note.** $E_0$ = kinetic fireball energy; $n$ = particle number density in the circumburst environment; $\theta_c$ = half-opening angle of the jet core; $\iota$ = viewing angle with respect to jet axis; $p$ = electron energy distribution power-law index; $M_{\rm ej}^{\rm dyn}$ = dynamical ejecta mass; $M_{\rm ej}^{\rm wind}$ = wind ejecta mass; $M_{\rm ej}$ = total ejecta mass; $v_{\rm ej}$ = ejecta velocity; $\Phi$ = half-opening angle of lanthanide-rich equatorial ejecta; $\epsilon_e$ = shock energy fraction that goes into the electrons; $\epsilon_B$ = shock energy fraction that goes into the magnetic energy density; $\chi_{\rm lan}$ = lanthanide mass fraction (D. Kasen et al. 2017); $\kappa_r$ = opacity; $\beta$ = power-law index of the ejecta mass distribution as a function of its velocity (B. D. Metzger 2017); and $\ln(Z)$ = natural logarithm of the Bayes evidence. We note that the Bu19-BNS-Jet model is the same as the one presented in Section 3.3 where now $\epsilon_e$ and $\epsilon_B$ are fixed.

## ORCID iDs

G. Stratta ● https://orcid.org/0000-0003-1055-7980
A. M. Nicuesa Guelbenzu ● https://orcid.org/0000-0002-6856-9813
S. Klose ● https://orcid.org/0000-0001-8413-7917
A. Rossi ● https://orcid.org/0000-0002-8860-6538
P. Singh ● https://orcid.org/0000-0003-1006-6970
E. Palazzi ● https://orcid.org/0000-0002-8691-7666
C. Guidorzi ● https://orcid.org/0000-0001-6869-0835
A. Camisasca ● https://orcid.org/0000-0002-4200-1947
S. Bernuzzi ● https://orcid.org/0000-0002-2334-0935
A. Rau ● https://orcid.org/0000-0001-5990-6243
M. Bulla ● https://orcid.org/0000-0002-8255-5127
F. Ragosta ● https://orcid.org/0000-0003-2132-3610
E. Maiorano ● https://orcid.org/0000-0003-2593-4355
D. Paris ● https://orcid.org/0000-0002-7409-8114